\documentclass[twocolumn,secnumarabic,preprintnumbers,amsmath,amssymb,floatfix]{revtex4}
\usepackage{graphicx}
\usepackage{subfigure} 
\usepackage{amssymb}

\begin{document}

\title{Dynamics of delay-coupled excitable neural systems}

\author {M. A. Dahlem$^1$}
\author {G. Hiller$^1$}
\author {A. Panchuk$^{1,2}$}
\author {E. Sch\"oll$^1$}

\affiliation{$^1$Institut f{\"u}r Theoretische Physik, Technische Universit{\"a}t Berlin,  Hardenbergstra{\ss}e 36, D-10623 Berlin, Germany}
\affiliation{$^2$Institute of Mathematics, National Academy of
  Sciences of Ukraine, Kiev, Ukraine}

\begin{abstract}
We study the nonlinear dynamics of two delay-coupled neural systems
each modelled by excitable dynamics of FitzHugh-Nagumo type and
demonstrate that bistability between the stable fixed point and limit
cycle oscillations occurs for sufficiently large delay times $\tau$
and coupling strength $C$. As the mechanism for these delay-induced
oscillations we identify a saddle-node bifurcation of limit cycles.

\end{abstract}

\maketitle

\section{Introduction}

The brain may be conceived as a dynamic network of coupled neurons
\cite{HAK06,WIL99,GER02}. These neurons are excitable units which can
emit spikes or bursts of electrical signals.  In order to describe the
complicated interaction between billions of neurons in large neural
networks,  the neurons are often lumped into highly connected sub-networks 
or synchronized sub-ensembles. Such neural populations are
usually spatially localized and contain both excitatory and inhibitory
neurons \cite{WIL72}.

The simplest model to display features of neural interaction consists 
of two coupled neural systems. Starting from this simplest network motif, 
larger networks can be built, and their effects may be studied. For example, 
starting from two interconnected reticular thalamic neurons with oscillatory 
behavior, it was shown in \cite{DES94} how more complex dynamics emerges in 
ring networks with nearest neighbors and fully reciprocal connectivity, or 
in networks organized in a two-dimensional array with proximal connectivity 
and ``dense proximal'' coupling in which every neuron connects to all other 
neurons within some radius.  In another example \cite{ZHO06c}, a neural population 
was itself modeled as a small sub-network of excitable elements, to study 
hierarchically clustered organization of excitable elements in a {\em network of networks}.

Most studies represent a population as an effective oscillatory element that is 
then coupled with other populations to study synchronization and network effects 
\cite{ROS04a,POP05,GAS07b}. In these studies, the basis for the emergence 
of complex network dynamics is the {\em oscillatory} behavior of neural elements, 
in spite of the fact that individual neural systems are usually in a stable steady 
state exhibiting {\em excitable} dynamics when perturbed.  A reason is that 
self-sustained oscillatory behavior in the individual element is required to define 
synchronization of subsystems \cite{ROS01a}.  One way to resolve this is to add weak 
Gaussian white noise to each individual element to generate sparse Poisson-like 
irregular spiking patterns, as seen in real neurons \cite{ZHO06c,HAU06,HOE07}.

In this report, we introduce a time-delayed coupling into the model of two excitable 
populations and demonstrate that the coupling-delay can induce sustained oscillations 
between the two subsystems.  We find a regime of bistability between the stable 
fixed point and limit cycle oscillations for sufficiently large delay times $\tau$ 
and coupling strength $C$. As the mechanism for these delay-induced oscillations 
we identify a saddle-node bifurcation of limit cycles. 

This result suggests the use of a compound system of two time-delayed coupled 
excitable elements as a minimal network motif to investigate  
oscillatory behavior in more complex networks.

\section{Model} 
We examine the delayed linear symmetric coupling of two
identical neural populations.  Each population is represented by a
simplified FitzHugh-Nagumo (FHN) system \cite{FIT61,NAG62}, which is
widely used as a paradigmatic model of excitable systems \cite{LIN04}.

The dynamical equations are given by:
  \begin{eqnarray}\label{eq:system}
    \epsilon \dot{x_1} &=& x_1 - \frac{x_1^3}{3} - y_1+ C[x_2(t-\tau)-x_1(t)]\nonumber\\
        \dot{y_1} &=& x_1 + a\nonumber\\
        \epsilon \dot{x_2} &=& x_2 - \frac{x_2^3}{3} - y_2 + C[x_1(t-\tau)-x_2(t)]\nonumber\\
        \dot{y_2} &=& x_2 + a
    \end{eqnarray}
where the two subsystems ($x_1,y_1$) and ($x_2,y_2$) correspond
to two neuron populations.  The value of the excitability parameter $a$ determines
whether the subsystem is excitable ($a>1$) or exhibits self-sustained
periodic firing ($a<1$).  This is because at $a\!=\!1$ the uncoupled systems
exhibit a Hopf bifurcation, and the fixed point becomes an unstable
focus for $a\!<\!1$.  $\epsilon$ is the timescale parameter that, if
chosen to be much smaller than unity, results in fast activator
variables $x_1,x_2$, and slow inhibitor variables $y_1,y_2$.  Unless otherwise
noted, we shall choose $\epsilon=0.01$ and $a\!=\!1.3$ for numerical
simulations and thus restrict our analysis to parameter values where
each of the two subsystems exhibits excitability with a stable fixed
point.

The interaction between the two neural populations is modelled as 
diffusive, i.\,e., the coupling vanishes if the variables $x_i$
are identical. The coupling strength $C$ is taken to be
symmetric for simplicity.  Further, we assume a finite signal propagation speed
between distant neural populations.  This is incorporated using the
delay time $\tau$.  Note that the coupling term has
the form of a classical diffusion term, but has lost its diffusive
character through the introduction of a propagation delay $\tau$.
More general delayed couplings are studied in \cite{BUR03}.

Before we analyse Eq.\ (\ref{eq:system}), let us briefly give 
a biophysical interpretation of the
hitherto abstract nature of the dynamical variables $x_i$ and $y_i$.  This
is needed to relate our results to other work on population dynamics
in neural networks, in particular to studies dealing with spiking rates
and time delays \cite{PIN01,COO03,HUT03,HUT04}.  These studies
consider {\em neural fields} \cite{WIL73,AMA77}, that is, they extend
over one or more spatial dimensions.  While neural field models adopt
the continuum limit of a network, we consider the minimal network
motif, i.\,e., two discrete subsystems, but in both cases the
biophysical reason to include delay is the same.

The generic biophysical interpretation of the FHN model is based on a
single point-like neuron and was originally not derived from features
of a neural population.  The variable $x$ models fast changes of the
electrical potential across the membrane (spikes), and $y$ is related
to the gating mechanism of membrane channels \cite{FIT61,NAG62}.  On
the contrary, a neural population is generically described by a time
coarse-grained mean-field model, in which the dynamical variables
represent averaged spiking rates \cite{WIL72}.  Notwithstanding, spiking
rates in neural populations can exhibit both steady state values and
relaxation oscillations. Both states are described by the FHN
mechanism with $a\!>\!1$ and $a\!<\!1$, respectively, which justifies
the FHN model for ``point-like'' populations.  To avoid confusion
about spikes vs.\ spiking rates, a model of uncoupled subsystems as
defined by Eq.~(\ref{eq:system}) with $C=0$ is sometimes referred to as
the Bonhoeffer-van der Pol model \cite{POL26,POL29,BON48}, for example
in Ref.\ \cite{ROS04}.


\section{Linear Stability Analysis}

The unique fixed point of the system is symmetric and is given by 
${\bf x}^* \equiv (x_1^*, y_1^*, x_2^*, y_2^*)$, where $x_i^*\!=\!-\!a$, $y_i^*\!=\!a^3/3\!-\!a$.

Denoting for convenience ${\bf x}(t\!-\!\tau) \equiv \tilde {\bf x}$, we can re-write 
system Eq.~(\ref{eq:system}) as: 
\begin{eqnarray}
\left(\begin{array}{c}\dot{x_1} \\\dot{y_1} \\\dot{x_2} \\\dot{y_2}\end{array}\right) = \left(\begin{array}{c}
f_1(\bf{x})\\f_2(\bf{x})\\f_3(\bf{x})\\f_4(\bf{x})\end{array}\right)+\left(\begin{array}{c}g_1(\bf{\tilde{x}})\\g_2(\bf{\tilde{x}})\\g_3(\bf{\tilde{x}})\\g_4(\bf{\tilde{x}})\end{array}\right)
\end{eqnarray}
This system can be linearized around the fixed point $\bf{x}^*$ by setting 
${\bf x}(t)=\bf{x}^*+\delta\bf{x}(t)$: 
\begin{eqnarray}
\label{eq:delaylinearized}
\delta\dot{\bf{x}} = \bf{J}_f^* \delta\bf{x} + \bf{\tilde{J}}_g^* \delta\tilde{\bf{x}}
\end{eqnarray}
with the Jacobian matrices $\bf{J}_f^*$ and $\bf{\tilde{J}}_g^*$.
The explicit form is
\begin{eqnarray}
\delta\dot{\bf{x}}
= \frac{1}{\epsilon} \left( \begin{array}{cccc}
\xi & -1 & 0 & 0\\
\epsilon & 0 & 0 & 0\\
0 & 0 & \xi & -1\\
0 & 0 & \epsilon & 0\end{array} \right)
\delta\bf{x} 
+ \frac{1}{\epsilon} \left( \begin{array}{cccc}
0 & 0 & C & 0\\
0 & 0 & 0 & 0\\
C & 0 & 0 & 0\\
0 & 0 & 0 & 0\end{array} \right)	
\delta\tilde{\bf{x}}
\end{eqnarray}
where $\xi=1-a^2-C$.
The ansatz
\begin{equation}\delta\bf{x}(t)=e^{\lambda t} \bf{u}\end{equation}
where $\bf{u}$ is an eigenvector of $\bf{J}_f^*$ implies
\begin{equation}\label{eq:ansatz}\delta\bf{\tilde x}=e^{\lambda t}e^{-\lambda \tau} \bf{u}\end{equation}
This leads to the characteristic equation 
\begin{equation}
(1- \xi \lambda+\epsilon \lambda^2 )^2-(\lambda C e^{-\lambda \tau })^2=0,
\end{equation}
which can be factorized giving 
\begin{equation}
1- \xi \lambda+\epsilon \lambda^2 =\pm \lambda C e^{-\lambda \tau }
\label{char_eq}
\end{equation}

\begin{figure}[t!]
\includegraphics[width=0.5\textwidth]{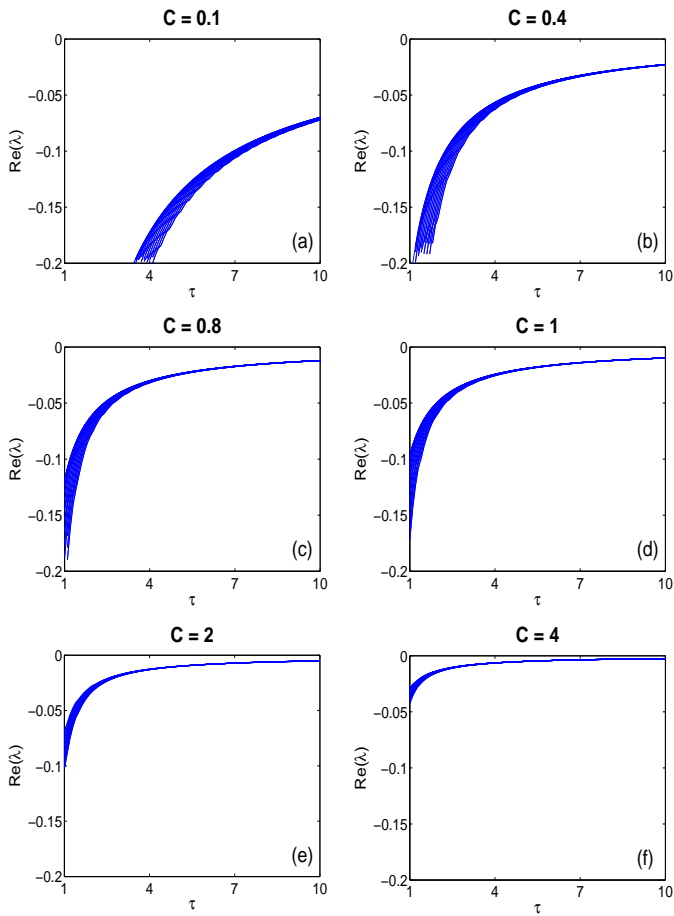} 
\caption{\label{fig:eigenmodes} Real parts Re($\lambda$) of the eigenvalues 
of the fixed point vs. time delay $\tau$ for $\epsilon=0.01$, $a=1.3$, 
and (a) C=0.1, (b) C=0.4, (c) C=0.8, (d) C=1, (e) C=2, (f) C=4.  
}
\end{figure}

This transcendental equation has infinitely many complex solutions $\lambda$. 
Fig.~\ref{fig:eigenmodes} shows the real parts of $\lambda$ for various values of $C$.
As can be seen in Fig.~\ref{fig:eigenmodes} the real parts of all eigenvalues are negative throughout, i.\,e., the fixed point of the coupled system remains stable for all $C$.  This can be shown analytically for $a>1$ by demonstrating that no delay-induced Hopf bifurcation can occur. Substituting the ansatz $\lambda=i \omega$ into Eq.~(\ref{char_eq}) and separating into real and imaginary parts yields for the imaginary part \begin{equation} \xi =\pm C cos(\omega \tau) \end{equation} This equation has no solution for $a>1$ since $|\xi|=a^2-1+C>C$, which proves that a Hopf bifurcation cannot occur.

\section{Delay-induced oscillations}
The cooperative dynamics of delay-induced oscillations in excitable systems 
is inherently different from those of noise-induced oscillations. The introduction of 
noise terms induces sustained oscillations in each individual subsystem by continuously 
kicking these subsystems out of their respective rest states.   Coupling then 
produces synchronisation effects between these individual oscillators \cite{HAU06,HOE07}.  
If time-delayed feedback control \cite{SCH07} is applied locally to one of the 
subsystems, the stochastic synchronisation can be tuned by varying the time-delay. 
This is in line with other work where it was demonstrated that such time-delayed 
feedback can be used to control the coherence and the timescales of noise-induced 
oscillations in a single FitzHugh-Nagumo system \cite{JAN04,BAL04,PRA07,JAN07}.

For {\em delayed} coupling the case is entirely different.  Here the sustained 
oscillations are an effect of the 
cooperative dynamics.  They are generated by the delayed interaction between two 
non-oscillating stable units, and are thus an {\em emergent} phenomenon of the 
compound system.  The bifurcation parameters for delay-induced bifurcations  are the 
coupling parameters $C$ and $\tau$.  

\begin{figure}[t!]
\includegraphics[width=0.5\textwidth]{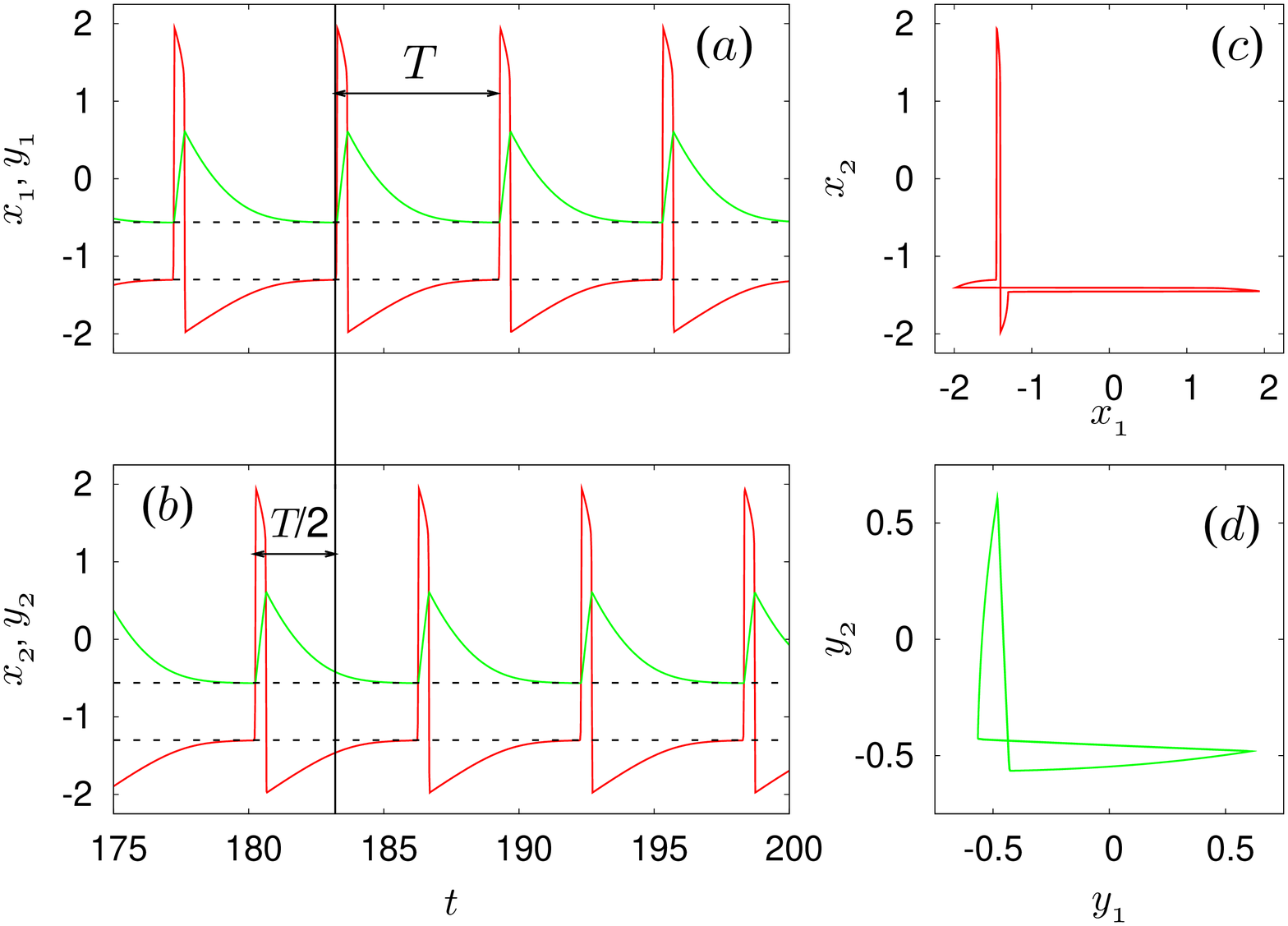} 
\caption{\label{fig:ts_comparison} Delay-induced oscillations. (a),
  (b): Time series of both subsystems (red solid lines: activator
  $x_i$, green solid lines: inhibitor $y_i$; black dashed lines: fixed
  point values of activator and inhibitor). (c), (d):
  Phase portraits of
  activators (c) and inhibitors (d).  Parameters: $\epsilon=0.01$,
  $a=1.3$, $C=0.5$, $\tau=3$. }
\end{figure}

For large coupling delay $\tau$ the oscillation is readily understood as the two 
units firing alternately, each spike initiated by the delayed signal of the remote 
system.  Since the signal of one system is transmitted to the other and then back, 
the oscillation in one system must have a period $T$ of approximately $2\tau$, 
and be phase shifted by $T/2$ with respect to the other system. This is visible in the 
time series (Fig.~\ref{fig:ts_comparison}(a),(b)) and in the phase portraits of the 
activators (Fig.~\ref{fig:ts_comparison}(c)) and inhibitors (Fig.~\ref{fig:ts_comparison} (d)). 

\begin{figure}[b!]
\includegraphics[width=0.5\textwidth]{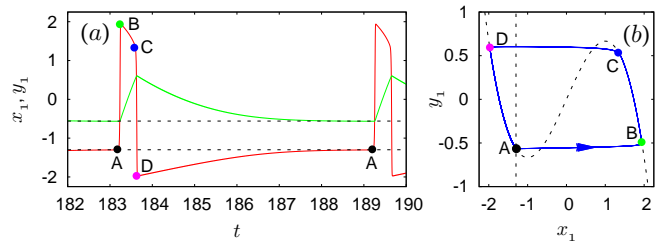} 
\caption{\label{fig:anatomy} 
Blow-up of delay-induced oscillation in the first subsystem: (a) time series and 
(b) phase portrait $(x_1,y_1)$. The four different stages of the limit cycle are 
separated by colored dots A, B, C, D.  Parameters: $\epsilon=0.01$, $a=1.3$, $C=0.5$, 
$\tau=3$.}
\end{figure}

\begin{figure}[b!]
\includegraphics[width=0.5\textwidth]{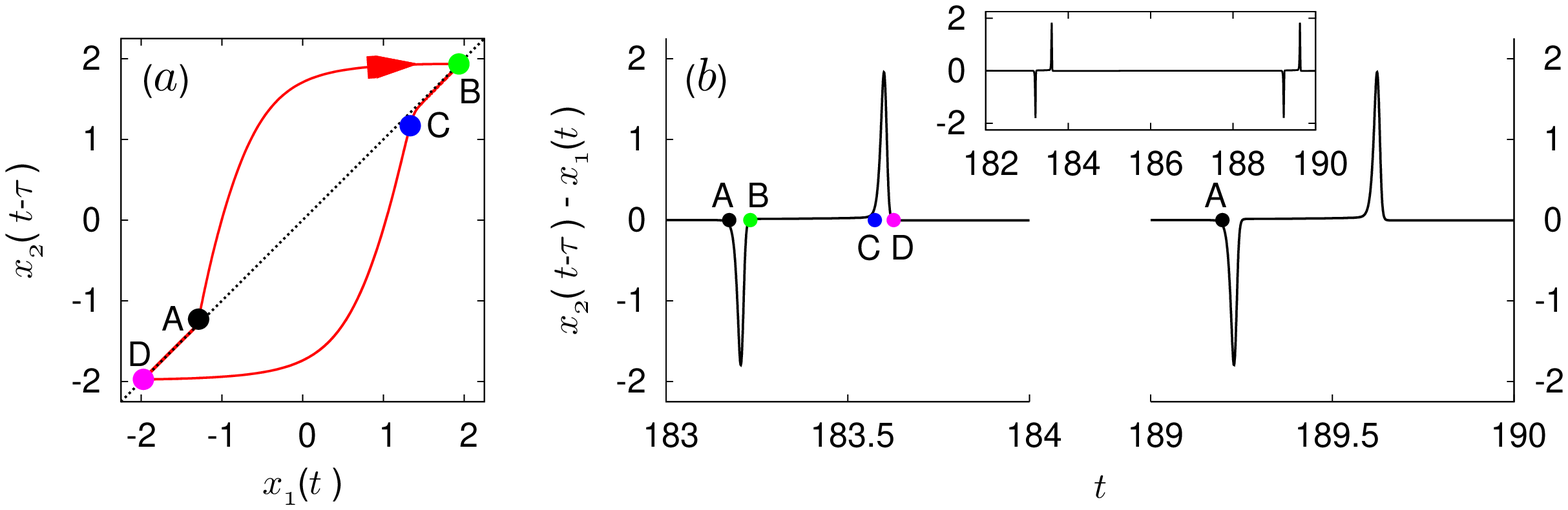} 
\caption{\label{fig:figSyncShifted}  
  (a) Phase portrait of delay-coupled excitable system in the plane
  \mbox{$(x_1(t),x_2(t-\tau))$} with positions on the orbit marked as
  colored dots $A, B, C, D$ as in Fig.~\ref{fig:anatomy}. (b) Time series of $(x_2(t-\tau)-x_1(t))$. 
  The inset shows a longer time series. 
  Parameters: $\epsilon=0.01$, $a=1.3$, $C=0.5$, $\tau=3$.}
\end{figure}

Since the two subsystems have identical but $T/2$-shifted profiles $x_i(t)$ and $y_i(t)$, the coupling terms \mbox{$C[x_i(t\!-\!\tau)\!-\!x_j(t)]$} ($i\neq j$) would vanish for all $t$ if the oscillation period $T$ were precisely $2\tau$.  Then the periodic orbit (Fig.~\ref{fig:anatomy}(b)) would also be a solution for the uncoupled subsystems ($C\!=\!0$). This cannot be the case for the excitable regime of the FHN systems ($a\!>\!1$) because periodic orbits do not exist there.  Hence it follows that the oscillation period $T$ must be of the form $T\!=\!2(\tau+\delta)$, where $\delta\!\neq\!0$ is the effective time shift responsible for the non-vanishing coupling term. The phase portraits of the activators $x_i$ in the planes \mbox{$(x_1(t),x_2(t-\tau))$} deviate from the bisector (Fig.~\ref{fig:figSyncShifted}(a)) due to the fact that $\delta\!\neq\!0$.  

To obtain a clear picture of the timescales involved in the dynamics, we have computed 
the excursion times along the segments of the phase-space trajectory. The start and 
end points of the different segments of the trajectory (colored dots $A$ to $D$) are 
defined to correspond to the time steps when the trajectory has left or entered a 
neighborhood $\Delta x_1$ (here $\Delta x_1\!=\!0.01$) of the stable branches of the 
$x_1$-nullcline.

\begin{figure}[t!]
\includegraphics[width=0.5\textwidth]{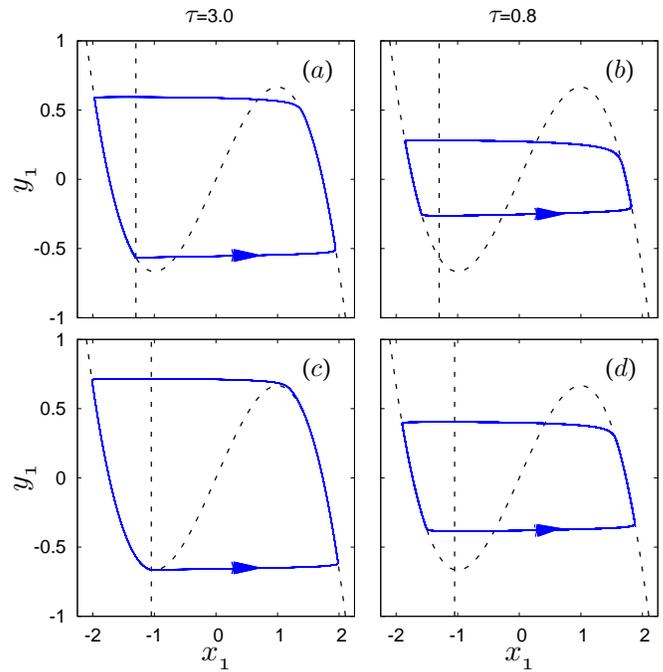} 
\caption{\label{fig:phaseportraits}  
Phase portraits of delay-coupled excitable system $(x_1,y_1)$  for different excitability 
parameters $a$ and delay times $\tau$ (trajectories: solid blue, nullclines: dashed black). 
(a) \mbox{$a\!=\!1.3$}, $\tau\!=\!3$, (b) $a\!=\!1.3$, $\tau\!=\!0.8$, (c) $a\!=\!1.05$, 
$\tau \!=\! 3$, (d) $a\!=\!1.05$, $\tau\!=\!0.8$.  Other parameters: $\epsilon\!=\!0.01$, 
$C\!=\!0.5$. }
\end{figure}

\begin{figure}[t!]
\includegraphics[width=0.5\textwidth]{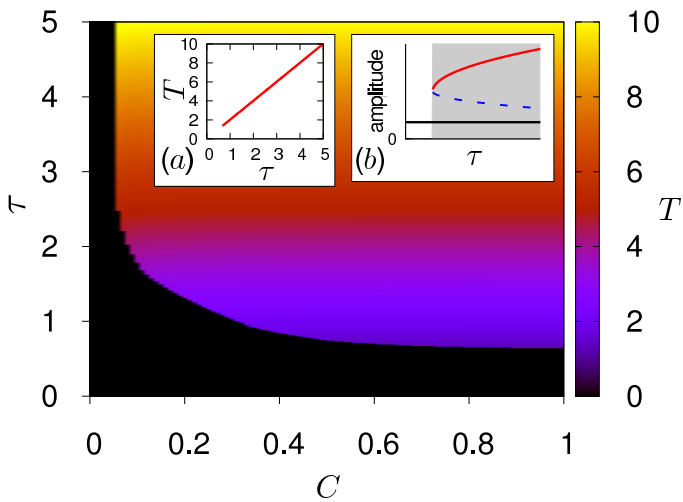} 
\caption{\label{fig:hauplot} Regime of oscillations in the $(\tau, C)$
  parameter plane for initial conditions corresponding to
  single-pulse-excitation in one system. The oscillation period $T$ is
  color coded. The transition between black and color marks the
  bifurcation line.  Inset (a) shows the oscillation period vs.\ time
  delay $\tau$ in a cut at $C\!=\!0.8$.  Parameters:
  $\epsilon\!=\!0.01$, $a\!=\!1.3$.  Inset (b): schematic plot of the
  saddle-node bifurcation of a stable (red solid line) and unstable
  (blue dashed) limit cycle. The maximal oscillation amplitude is
  plotted vs. the delay time $\tau$ and the stable fixed point is
  plotted as a solid black line. The gray background marks the
  bistable region.}
\end{figure} 

\begin{figure}[b!]
\includegraphics[width=0.5\textwidth]{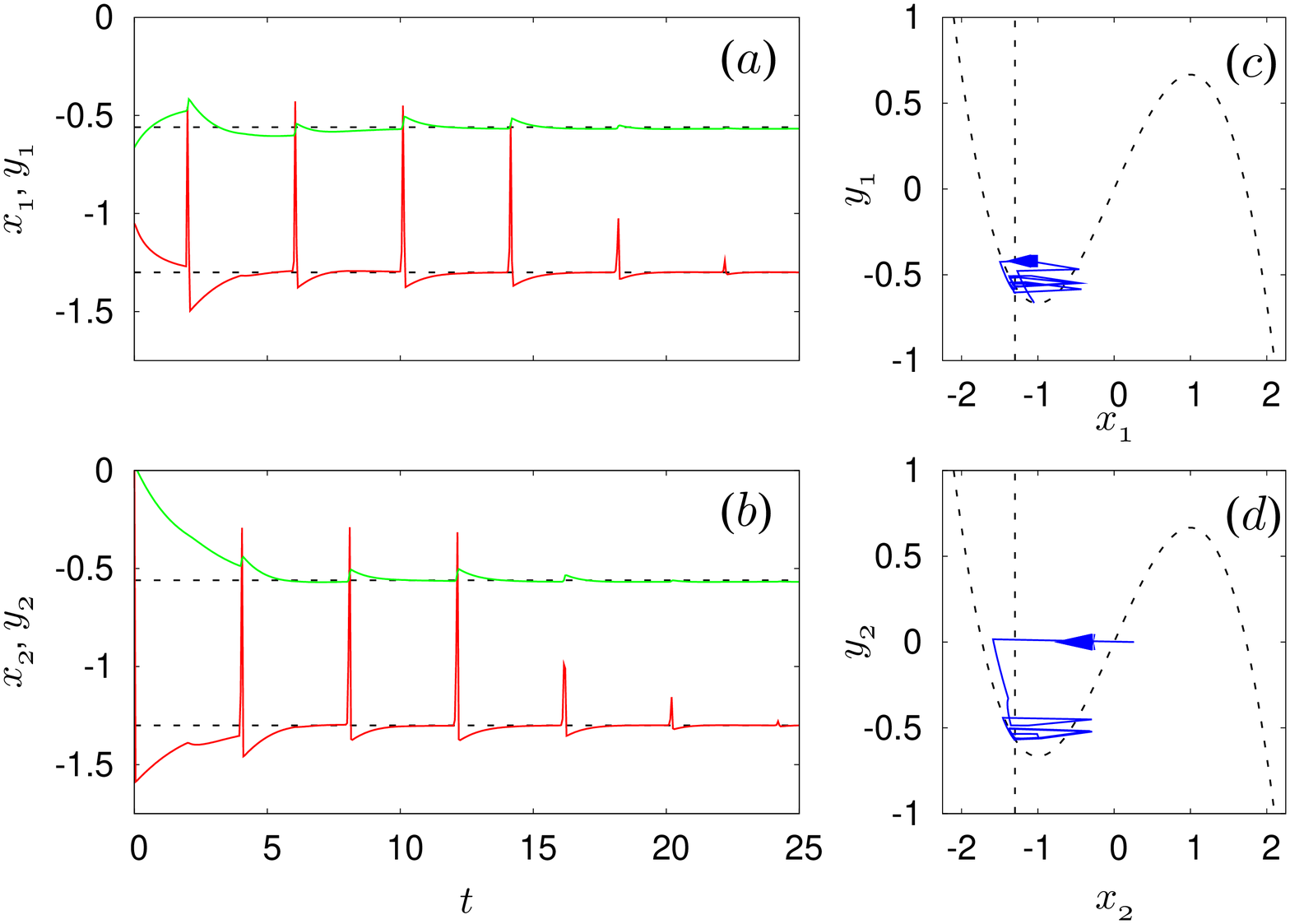} 
\caption{\label{fig:transients} 
Unstable delay-induced limit cycle. (a), (b): 
Time series of both subsystems (red solid lines: activator $x_i$, green solid
lines: inhibitor $y_i$; black dashed lines: fixed point value of activator and
inhibitor). (c),(d): phase portraits of $(x_1\!,\!y_1)$ (top) and $(x_2\!,\!y_2)$ (bottom).
The unstable small-amplitude limit cycle is visualized by the initial part
of the transient approaching the stable fixed point.
Parameters: $\epsilon\!=\!0.01$, $a\!=\!1.3$, $C\!=\!0.5$,  $\tau\!=\!3$.}
\end{figure}

In the example shown in Fig.~\ref{fig:anatomy} 
($\epsilon\!=\!0.01$, $a\!=\!1.3$, $C\!=\!0.5$, $\tau\!=\!3$) the value of the effective 
time shift is $\delta\!=\!0.012$.  This value is about one third of the fast transition 
times between the stable branches of the $x_1$-nullcline, $\Delta t_1\!=\!0.041$ and 
$\Delta t_3\!=\!0.038$, respectively.  $\Delta t_1$ is the rise-time of the  spike, 
i.e., the time that elapses between leaving the fixed point (black dot $A$) and crossing the 
right stable branch of the $x_1$-nullcline at $B$ (green dot, Fig.~\ref{fig:anatomy}(b)). 
$\Delta t_3$ is the drop-time, i.e., the duration of the jump back from the right to the 
left stable branch of the $x_1$-nullcline (between blue and magenta dots, $C \to D$). 
The slow parts of the trajectory, occurring on the right ($B \to C$) and left ($D \to A$) 
stable branches of the $x_1$-nullcline, have a duration $\Delta t_2\!=\!0.357$ and 
$\Delta t_4\!=\!5.588$, respectively. The total oscillation period is thus $T\!=\!6.024$.

In Fig.~\ref{fig:phaseportraits} the phase portraits of the first of the two 
delay-coupled subsystems are shown for different excitability parameters $a$ and delay
times $\tau$.  The top panel (a,b) corresponds to excitability parameters far from the 
Hopf bifurcation of the uncoupled system, which occurs at $a\!=\!1$.  The bottom panel 
(c,d) corresponds to values of $a$ close to the Hopf bifurcation.

In the case of $a\!=\!1.3$ and $\tau=3$ (Fig.~\ref{fig:phaseportraits}(a)), the 
oscillation period $T\!=\!6.024$ is large enough for the two subsystems to nearly 
approach the fixed point before being perturbed again by the remote signal.
Note that Fig.~\ref{fig:phaseportraits}(a) corresponds to
Fig.~\ref{fig:anatomy}(b).
If the delay time becomes much smaller, e.\,g., for $\tau\!=\!0.8$ 
(Fig.~\ref{fig:phaseportraits} (b)), the excitatory spike of the other subsystem arrives 
while the first system is still in the refractory phase, so that it cannot complete the 
return $D \to A$ to the fixed point. The times spent in the different stages of the limit 
cycle for $\tau\!=\!0.8$ are $\Delta t_1\!=\!0.087$, $\Delta t_2\!=\!0.111$, 
$\Delta t_3\!=\!0.085$, and $\Delta t_4\!=\!1.355$, and the oscillation period is 
$T\!=\!1.637$.  The effective time shift $\delta\!=\!0.018$ for $\tau\!=\!0.8$ is larger 
than for $\tau\!=\!3$.

We now investigate the transition from large to small delay times
$\tau$ close to the Hopf bifurcation at $a\!=\!1.05$.  Again we choose
$\tau\!=\!3$ (Fig.~\ref{fig:phaseportraits}(c)) and find $T\!=\!6.018$
and $\delta\!=\!0.009$.  The times spent in the different phases of
the oscillation period $T$ are $\Delta t_1\!=\!0.043$, $\Delta
t_2\!=\!0.329$, $\Delta t_3\!=\!0.186$, and $\Delta t_4\!=\!5.461$.
For the shorter delay time $\tau\!=\!0.8$
(Fig.~\ref{fig:phaseportraits}(d)) we find $T\!=\!1.630$,
$\delta\!=\!0.015$, $\Delta t_1\!=\!0.072$, $\Delta t_2\!=\!0.240$,
$\Delta t_3\!=\!0.070$, and $\Delta t_4\!=\!1.249$. We find the same
pattern as far from the Hopf bifurcation.  The effective time shift
$\delta$ is smaller if $\tau$ becomes larger and the system can
approach the fixed point. Furthermore, if the system is close to the
Hopf bifurcation and $\tau$ is sufficiently large so that the fixed
point is very closely approached, $\delta$ becomes much smaller than
far from the Hopf bifurcation, and tends to zero for $a \to 1$. 
  The reason is that when the excitatory spike arrives in the first
  subsystem, there is a turn-on delay $\delta$ before the first
  subsystem emits a spike. This is because the trajectory has to 
  cross the middle branch of the $x_1$-nullcline due to this {\em upstream} impulse. 
  This section of the trajectory becomes the
  smaller, the closer the fixed point $A$ (at $x_1\!=\!-a$) is to the
  minimum of the $x_1$ nullcline (at $x_1\!=\!-1$). This also explains
  the origin of the time-shift $\delta$.


Finally, we shall investigate the question whether the system exhibits bistability
between the fixed point and the limit cycle oscillation for all values of $\tau$. 
In Fig.~\ref{fig:hauplot} 
the regime of oscillations is shown
in the parameter plane of the coupling strength $C$ and coupling delay $\tau$. The
oscillation period is color coded. The boundary of this colored region is given by
the minimum coupling delay $\tau_{min}$ as a function of $C$. For large coupling strength
$\tau_{min}$ is almost independent of $C$; with decreasing $C$ it sharply increases, and
at some minimum $C$ no oscillations exist at all. 
At the boundary, the oscillation sets in with finite frequency and amplitude as can be 
seen in the inset of Fig.~\ref{fig:hauplot} which shows a cut of the parameter plane at $C\!=\!0.8$.
The oscillation period increases linearly with $\tau$.  The mechanism that 
generates the oscillation is a saddle-node bifurcation of limit cycles 
(see Fig.~\ref{fig:hauplot} inset (b)), creating a pair of a stable and an unstable limit cycle. 
The unstable limit cycle can be visualized from transients starting at 
appropriate initial conditions (see Fig.~\ref{fig:transients}). The trajectory
initially remains near the unstable limit cycle, which separates the two attractor
basins of the stable limit cycle and the stable fixed point, and then asymptotically 
approaches the stable fixed point.

\section{Conclusion}

We have shown that delayed coupling can induce periodic spiking in a 
compound system of two coupled neuron populations if the delay and the coupling strength are sufficiently 
large. Bistability of a fixed point and limit cycle oscillations occur even though the 
single excitable element displays only a stable fixed point. The two neural populations 
oscillate with a phase lag of $\pi$.

\section{Acknowledgements}
This work was supported by DFG in the framework of Sfb 555. The authors would like to 
thank P. H{\"o}vel, V. Flunkert, S. Brandstetter, J. Hofmann and F. Schneider for fruitful discussions.


\begin{thebibliography}{10}
\expandafter\ifx\csname url\endcsname\relax
  \def\url#1{{\tt #1}}\fi
\expandafter\ifx\csname urlprefix\endcsname\relax\def\urlprefix{URL }\fi

\bibitem{HAK06}
H.~Haken: {\em Brain Dynamics: Synchronization and Activity Patterns in
  Pulse-Coupled Neural Nets with Delays and Noise\/} (Springer Verlag GmbH,
  Berlin, 2006).

\bibitem{WIL99}
H.~R. Wilson: {\em Spikes, Decisions, and Actions: The Dynamical Foundations of
  Neuroscience\/} (Oxford University Press, Oxford, 1999).

\bibitem{GER02}
W.~Gerstner and W.~Kistler: {\em Spiking neuron models\/} (Cambridge University
  Press, Cambridge, 2002).

\bibitem{WIL72}
H.~R. Wilson and J.~D. Cowan: {\em Excitatory and inhibitory interactions in
  localized populations of model neurons.\/}, Biophysical journal {\bf 12}, 1
  (1972).

\bibitem{DES94}
A.~Destexhe, D.~Contreras, T.~J. Sejnowski, and M.~Steriade: {\em {{A} model of
  spindle rhythmicity in the isolated thalamic reticular nucleus}\/}, J.
  Neurophysiol. {\bf 72}, 803 (1994).

\bibitem{ZHO06c}
C.~Zhou, L.~Zemanova, G.~Zamora, C.~C. Hilgetag, and J.~Kurths: {\em
  Hierarchical organization unveiled by functional connectivity in complex
  brain networks\/}, Phys. Rev. Lett. {\bf 97}, 238103 (2006).

\bibitem{ROS04a}
M.~G. Rosenblum and A.~Pikovsky: {\em Controlling synchronization in an
  ensemble of globally coupled oscillators\/}, Phys.~Rev.~Lett. {\bf 92},
  114102 (2004).

\bibitem{POP05}
O.~V. Popovych, C.~Hauptmann, and P.~A. Tass: {\em Effective desynchronization
  by nonlinear delayed feedback\/}, Phys.~Rev.~Lett. {\bf 94}, 164102 (2005).

\bibitem{GAS07b}
M.~Gassel, E.~Glatt, and F.~Kaiser: {\em Time-delayed feedback in a net of
  neural elements: Transitions from oscillatory to excitable dynamics\/},
  Fluct. Noise Lett. {\bf 7}, L225 (2007).

\bibitem{ROS01a}
M.~G. Rosenblum, A.~Pikovsky, and J.~Kurths: {\em Synchronization -- A
  universal concept in nonlinear sciences\/} (Cambridge University Press,
  Cambridge, 2001).

\bibitem{HAU06}
B.~Hauschildt, N.~B. Janson, A.~G. Balanov, and E.~Sch{\"o}ll: {\em
  Noise-induced cooperative dynamics and its control in coupled neuron
  models\/}, Phys.~Rev.~E {\bf 74}, 051906 (2006).

\bibitem{HOE07}
P.~H{\"o}vel, M.~A. Dahlem, and E.~Sch{\"o}ll: {\em Synchronization of
  noise-induced oscillations by time-delayed feedback\/}, in {\em Proc. 19th
  Internat. Conf. on Noise and Fluctuations ({ICNF-2007})\/} (American
  Institute of Physics, College Park, Maryland 20740-3843, 2007).

\bibitem{FIT61}
R.~FitzHugh: {\em Impulses and physiological states in theoretical models of
  nerve membrane\/}, Biophys. J. {\bf 1}, 445 (1961).

\bibitem{NAG62}
J.~Nagumo, S.~Arimoto, and S.~Yoshizawa.: {\em An active pulse transmission
  line simulating nerve axon.\/}, Proc. IRE {\bf 50}, 2061 (1962).

\bibitem{LIN04}
B.~Lindner, J.~Garc{\'i}a-Ojalvo, A.~Neiman, and L.~Schimansky-Geier: {\em
  Effects of noise in excitable systems\/}, Phys.~Rep. {\bf 392}, 321 (2004).

\bibitem{BUR03}
N.~Buric and D.~Todorovic: {\em Dynamics of fitzhugh-nagumo excitable systems
  with delayed coupling\/}, Phys. Rev. E {\bf 67}, 066222 (2003).

\bibitem{PIN01}
D.~J. Pinto and G.~B. Ermentrout: {\em Spatially structured activity in
  synaptically coupled neuronal networks: I. traveling fronts and pulses\/},
  SIAM Journal on Applied Mathematics {\bf 62}, 206 (2001).

\bibitem{COO03}
S.~Coombes, G.~J. Lord, and M.~R. Owen: {\em Waves and bumps in neuronal
  networks with axo-dendritic synaptic interactions\/}, Physica D: Nonlinear
  Phenomena {\bf 178}, 219 (2003).

\bibitem{HUT03}
A.~Hutt, M.~Bestehorn, and T.~Wennekers: {\em Pattern formation in
  intracortical neuronal fields\/}, Network: Computation in Neural Systems {\bf
  14}, 351 (2003).

\bibitem{HUT04}
A.~Hutt: {\em Effects of nonlocal feedback on traveling fronts in neural fields
  subject to transmission delay\/}, Phys. Rev. E {\bf 70}, 052902 (2004).

\bibitem{WIL73}
H.~R. Wilson and J.~D. Cowan: {\em A mathematical theory of the functional
  dynamics of cortical and thalamic nervous tissue.\/}, Kybernetik {\bf 13}, 55
  (1973).

\bibitem{AMA77}
S.~Amari: {\em {{D}ynamics of pattern formation in lateral-inhibition type
  neural fields}\/}, Biol. Cybern. {\bf 27}, 77 (1977).

\bibitem{POL26}
B.~van~der Pol: {\em On relaxation oscillations\/}, Phil. Mag. {\bf 2}, 978
  (1926).

\bibitem{POL29}
B.~van~der Pol and J.~van~der Mark: {\em The heart beat considered as a
  relaxation oscillation and an electrical model of the heart\/}, Arch. Neerl.
  Physiol. {\bf 14}, 418 (1929).

\bibitem{BON48}
K.~F. Bonhoeffer: {\em {Activation of passive iron as a model for the
  excitation of nerve}\/}, J. Gen. Physiol. {\bf 32}, 69 (1948).

\bibitem{ROS04}
M.~G. Rosenblum and A.~Pikovsky: {\em Delayed feedback control of collective
  synchrony: An approach to suppression of pathological brain rhythms\/},
  Phys.~Rev.~E {\bf 70}, 041904 (2004).

\bibitem{SCH07}
E.~Sch{\"o}ll and H.~G. Schuster (Editors): {\em Handbook of Chaos Control\/}
  (Wiley-VCH, Weinheim, 2008).

\bibitem{JAN04}
N.~B. Janson, A.~G. Balanov, and E.~Sch{\"o}ll: {\em Delayed feedback as a
  means of control of noise-induced motion\/}, Phys.~Rev.~Lett. {\bf 93},
  010601 (2004).

\bibitem{BAL04}
A.~G. Balanov, N.~B. Janson, and E.~Sch{\"o}ll: {\em Control of noise-induced
  oscillations by delayed feedback\/}, Physica~D {\bf 199}, 1 (2004).

\bibitem{PRA07}
T.~Prager, H.~P. Lerch, L.~Schimansky-Geier, and E.~Sch{\"o}ll: {\em Increase
  of coherence in excitable systems by delayed feedback\/}, J. Phys. A {\bf
  40}, 11045 (2007).

\bibitem{JAN07}
N.~B. Janson, A.~G. Balanov, and E.~Sch{\"o}ll: {\em Control of noise-induced
  dynamics\/}, in {\em Handbook of Chaos Control\/}, edited by E.~Sch{\"o}ll
  and H.~G. Schuster (Wiley-VCH, Weinheim, 2008).

\end{thebibliography}

\end{document}